\documentclass[12pt,preprint]{emulateapj}
\usepackage{graphicx}

\shorttitle{The HD 38529 Planetary System}
\shortauthors{Gregory W. Henry et al.}
\slugcomment{Submitted for publication in the Astrophysical Journal}
\begin{document}

\title{Host Star Properties and Transit Exclusion for the HD 38529
  Planetary System}

\author{
  Gregory W. Henry\altaffilmark{1},
  Stephen R. Kane\altaffilmark{2},
  Sharon X. Wang\altaffilmark{3,4},
  Jason T. Wright\altaffilmark{3,4},
  Tabetha S. Boyajian\altaffilmark{5},
  Kaspar von Braun\altaffilmark{2,6},
  David R. Ciardi\altaffilmark{2},
  Diana Dragomir\altaffilmark{7},
  Chris Farrington\altaffilmark{8},
  Debra A. Fischer\altaffilmark{5},
  Natalie R. Hinkel\altaffilmark{2},
  Andrew W. Howard\altaffilmark{9},
  Eric Jensen\altaffilmark{10},
  Gregory Laughlin\altaffilmark{11},
  Suvrath Mahadevan\altaffilmark{3,4},
  Genady Pilyavsky\altaffilmark{3}
}
\email{gregory.w.henry@gmail.com}
\altaffiltext{1}{Center of Excellence in Information Systems, Tennessee
  State University, 3500 John A. Merritt Blvd., Box 9501, Nashville,
  TN 37209}
\altaffiltext{2}{NASA Exoplanet Science Institute, Caltech, MS 100-22,
  770 South Wilson Avenue, Pasadena, CA 91125}
\altaffiltext{3}{Department of Astronomy and Astrophysics,
  Pennsylvania State University, 525 Davey Laboratory, University
  Park, PA 16802}
\altaffiltext{4}{Center for Exoplanets \& Habitable Worlds,
  Pennsylvania State University, 525 Davey Laboratory, University
  Park, PA 16802}
\altaffiltext{5}{Department of Astronomy, Yale University, New Haven,
  CT 06511}
\altaffiltext{6}{Max Planck Institut for Astronomy, K\"{o}nigstuhl 17,
  69117, Heidelberg, Germany}
\altaffiltext{7}{Las Cumbres Observatory Global Telescope Network,
  6740B Cortona Dr., Suite 102, Goleta, CA 93117, USA}
\altaffiltext{8}{The CHARA Array, Mount Wilson Observatory, Mount
  Wilson, CA 91023, USA}
\altaffiltext{9}{Institute for Astronomy, University of Hawaii,
  Honolulu, HI 96822, USA}
\altaffiltext{10}{Dept of Physics \& Astronomy, Swarthmore College,
  Swarthmore, PA 19081}
\altaffiltext{11}{UCO/Lick Observatory, University of California, Santa
  Cruz, CA 95064}

%%%%%%%%%%%%%%%%%%%%%%%%%%%%%%%%%%%%%%%%%%%%%%%%%%%%%%%%%%%%%%%%%%%%

\begin{abstract}

The transit signature of exoplanets provides an avenue through which 
characterization of exoplanetary properties may be undertaken, such as 
studies of mean density, structure, and atmospheric composition. The 
Transit Ephemeris Refinement and Monitoring Survey (TERMS) is a program to 
expand the catalog of transiting planets around bright host stars by 
refining the orbits of known planets discovered with the radial velocity 
technique. Here we present results for the HD~38529 system. We determine 
fundamental properties of the host star through direct interferometric 
measurements of the radius and through spectroscopic analysis. We provide 
new radial velocity measurements that are used to improve the Keplerian 
solution for the two known planets, and we find no evidence for a previously 
postulated third planet. We also present 12 years of precision robotic 
photometry of HD~38529 that demonstrate the inner planet does not transit 
and the host star exhibits cyclic variations in seasonal mean brightness 
with a timescale of approximately six years.

\end{abstract}

\keywords{planetary systems -- techniques: photometric -- techniques:
  radial velocities -- stars: individual (HD~38529)}

%%%%%%%%%%%%%%%%%%%%%%%%%%%%%%%%%%%%%%%%%%%%%%%%%%%%%%%%%%%%%%%%%%%%

\section{Introduction}
\label{introduction}

The study of transiting exoplanets has undergone a remarkable
evolution since the first transit detection in HD~209458b
\citep{cha00,hen00}.  Projects such as the Hungarian Automated
Telescope Network (HATNet) \citep{bak04}, and SuperWASP \citep{pol06}
are routinely detecting new transiting planets. Results from the NASA
Kepler mission are breaking new ground by discovering planets that are
smaller and at longer orbital periods than those accessible from
ground-based surveys \citep{bor11a,bor11b}. However, the most
important transiting planets for follow-up observations and
characterization of atmospheres continue to be those discovered first
by the radial velocity (RV) method, because of the bias towards bright
host stars. The Transit Ephemeris Refinement and Monitoring Survey
(TERMS) seeks to provide more of these opportunites through orbital
refinement \citep{kan09}, particularly for planets in eccentric orbits
with higher transit probabilities \citep{kan08}. TERMS also allows the
study of long-term variability of the host star such that it can be
correlated with the stellar magnetic activity cycle in the context of
being a planet-host \citep{dra12}.
  
Multi-planet systems present particularly interesting cases since
studying the dynamical interaction of the planets is an additional
advantage to refining the orbits. One such TERMS target is the
planetary system orbiting HD~38529, whose planets have been discovered
via precision RV measurements.  The inner planet is in a $\sim 14$~day
orbit and was discovered by \citet{fis01}. The second planet, in a
$\sim 6$~year orbit, was discovered by \citet{fis03}. The system was
further investigated in the context of multi-planet systems by
\citet{wit09} and \citet{wri09b}, both of whom provided new RV data
and revised orbits for the planets. \citet{ben10} furthered the
studies of the system by providing new RV data as well as Hubble Space
Telescope astrometry. Their analysis indicated that there may be an
additional planet located at an orbital period of $\sim194$~days,
though this was not confirmed by their observations and they
encouraged further study to help resolve the issue.

\begin{deluxetable*}{rccc}
\tablecaption{Stellar Properties of HD~38529 \label{stellar}} 
\tablewidth{0pc}
\tablehead{
\colhead{ } &
\colhead{Value} &
\colhead{Value}	&
\colhead{ }	\\
\colhead{Parameter}	&
\colhead{Spectroscopic}	&
\colhead{Interferometric} &
\colhead{Reference}		
}
\startdata
$\theta_{\rm UD}$ (mas) \dotfill		&	\nodata	&	$0.593\pm0.016$	&	this work (\S \ref{sec:stellar_radius})	\\
$\theta_{\rm LD}$ (mas)	\dotfill	&		\nodata	&	$0.611\pm0.016$	&	this work (\S \ref{sec:stellar_radius})	\\
Luminosity (L$_{\rm \odot}$) \dotfill	& \nodata 	&	$5.777 \pm 0.186$	&	this work (\S \ref{sec:effective_temperature})	\\
Radius $R_*$ (R$_{\rm \odot}$) \dotfill	&	$2.34 \pm 0.07$		&	$2.578 \pm 0.080$ 	&	this work (\S \ref{sme}, \S \ref{sec:stellar_radius})\\
$T_{\rm eff}$ (K)	\dotfill			&	$5619 \pm 44$\phn\phn	&	$5576 \pm 74$\phn\phn		&	this work (\S \ref{sme}, \S \ref{sec:effective_temperature}) \\ 
$[$Fe/H$]$ \dotfill		&	$0.38 \pm 0.03$	&	\nodata	&	this work (\S \ref{sme})	\\
$v \sin i$ (km~s$^{-1}$)	\dotfill	&	$3.20 \pm 0.50$	&	\nodata	&	this work (\S \ref{sme})	\\
$\log g$ \dotfill	&	$3.83 \pm 0.06$	&	\nodata 	&	 this work (\S \ref{sme})	\\
Mass $M_*$ (M$_{\rm \odot}$) \dotfill	&	$1.36 \pm 0.02$		&	\nodata 	&	this work (\S \ref{sme})	\\
Age (Gyr)	\dotfill	& $4.45 \pm 0.23$	&	\nodata	&	this work (\S \ref{sme}) \\ 
\enddata
\tablecomments{For details, see \S \ref{sec:fundamental_parameters} and \S \ref{sme}.}
\end{deluxetable*}

In this paper we present an exhaustive analysis of both the host star
and the planets in the HD~38529 system. We performed interferometric
observations of the host star using the CHARA Array to determine a
direct measurement of the radius, which we compare with that derived
from an analysis of high-resolution spectra. We provide new RV data
for the system, improving the orbits of the planets and greatly
extending the time baseline. Our analysis of the combined data finds
no evidence for a third planet located near 194 days. From these new
data, we calculate an accurate transit ephemeris and predict the
parameters of a potential transit of HD~38529b.  We present 12 years
of precision photometry of HD~38529 acquired with an automated
photoelectric telescope (APT) at Fairborn Observatory. These
observations rule out a transit of the inner planet. Furthermore, we
demonstrate that the host star exhibits a cyclic brightness variation
on a timescale of approximately six years which is correlated with the
S-index derived from our Keck I spectra.

%%%%%%%%%%%%%%%%%%%%%%%%%%%%%%%%%%%%%%%%%%%%%%%%%%%%%%%%%%%%%%%%%%%%

\section{Fundamental Stellar Parameters}
\label{sec:fundamental_parameters}

%%%%%%%%%%%%%%%%%%%%%%%%%%%%%%%%%%%%%%%%%%%%%%%%%%%%%%%%%%%%%%%%%%%%

\subsection{Stellar Radius}
\label{sec:stellar_radius}

HD~38529 was observed during three nights in November 2012 using the 
Georgia State University Center for High Angular Resolution Astronomy 
(CHARA) interferometric array \citep{ten05}. Our observational methods 
and strategy are described in detail in \citet{von12} and \citet{boy12b} 
and references therein. We used two of CHARA's longest baselines (S1E1 
and E1W1) to perform our observations in $H$-band with the CHARA Classic 
beam combiner \citep{stu03,ten05} in single-baseline mode. We obtained 3, 
1, and 2 observations (brackets) during the nights of 3, 4, and 12 November 
2012, each of which contains approximately 2.5 minutes of integration and 
1.5 minutes of telescope slewing per object (target and calibrator). To 
remove the influence of atmospheric and instrumental systematics, 
interferometric observations consist of bracketed sequences of object and 
calibrator stars, chosen to be near-point-like sources of similar brightness 
as HD~38529 and located at small angular distances from it. We originally 
used both HD~37077 and HD~36777 as calibrators but eliminated the latter 
due to the presence of a second fringe packet in each observation, 
indicating that HD~36777 may be an unresolved binary.

To protect against unknown systematics in interferometric data, we ordinarily 
require the use of at least two calibrators, two baselines, and data obtained 
during at least two nights \citep{boy12a,boy12b,von11a,von11b,von12}. Thus, 
we added 12 archival CHARA $K$-band brackets obtained in 2005 using the 
S1E1 baseline and HD~43318 as calibrator, published by \citet{bai08}.

\begin{figure}
  \includegraphics[width=8.2cm]{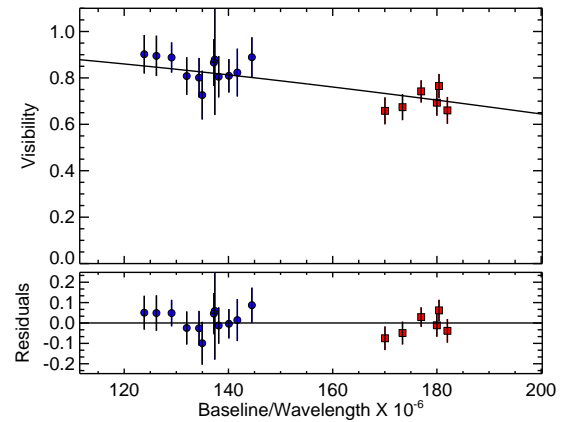}
  \caption{Calibrated visibility observations along with the limb-darkened 
    angular diameter fit for HD~38529 (top panel) along with the fractional 
    residuals around the fit (bottom panel). The blue round points are the 
    $K$-band data, and the red squares are the $H$-band data. The $K$-band 
    data from \citet{bai08} were taken with a shorter effective baseline than 
    our new $H$-band data (due to the difference in wavelength). The reduced 
    $\chi^2$ of this fit is 0.49. For more details, see 
    \S \ref{sec:stellar_radius}.}
  \label{fig:diameters}
\end{figure}

The uniform disk and limb-darkened angular diameters ($\theta_{\rm
  UD}$ and $\theta_{\rm LD}$, respectively; see Table \ref{stellar})
are found by fitting the calibrated visibility measurements (Figure
\ref{fig:diameters}) to the respective functions for each relation.
These functions may be described as $n^{th}$-order Bessel functions of
the angular diameter of the star, the projected distance between the
two telescopes, and the wavelength of observation (see equations 2 and
4 of \citet{han74}). Visibility is the normalized amplitude of the
correlation of the light from two telescopes.  It is a unitless number
ranging from 0 to 1, where 0 implies no correlation and 1 implies
perfect correlation. An unresolved source would have a perfect
correlation of 1.0 independent of the distance between the telescopes
(baseline). A resolved object will show a decrease in visibility with
increasing baseline length. The shape of the visibility versus
baseline is a function of the topology of the observed object (the
Fourier Transform of the object's shape). For a uniform disk this
function is a Bessel function, and for this paper, we use a simple
model for limb-darkening variation of a uniform disk. The visibility
of any source is reduced by a non-perfect interferometer, and the
point-like calibrators are needed to calibrate out the loss of
coherence caused by instrumental effects.

We use the linear limb-darkening coefficient $\mu_{H} = 0.362$ from the 
ATLAS models in \citet{cla00} for stellar $T_\mathrm{eff}$ = 5500~K and 
$\log g$ = 4.0 to convert from $\theta_{\rm UD}$ to $\theta_{\rm LD}$. 
The uncertainties in the adopted limb darkening coefficient amount to 
0.2\% when modifying the adopted gravity by 0.5~dex or the adopted 
$T_\mathrm{eff}$ by 200K, well within the errors of our diameter estimate.

Our interferometric measurements yield the following values for HD~38529's 
angular diameters: $\theta_{\rm UD} = 0.593 \pm 0.016$~milliarcseconds (mas) 
and $\theta_{\rm LD} = 0.611 \pm 0.016$~mas (Table~\ref{stellar}). Combined 
with the direct distance measurement from \citet{van07} of 
$39.277 \pm 0.617$ pc, we derive a stellar radius for HD~38529 of 
$2.5780 \pm 0.0795 R_{\odot}$, which is consistent with 
$2.44 \pm 0.22 R_{\odot}$ calculated by \citet{bai08}.

%%%%%%%%%%%%%%%%%%%%%%%%%%%%%%%%%%%%%%%%%%%%%%%%%%%%%%%%%%%%%%%%%%%%

\subsection{Stellar Effective Temperature and Luminosity}
\label{sec:effective_temperature}

To calculate HD~38529's effective temperature and luminosity, we
produce a spectral energy distribution (SED) fit based on the spectral
templates in \citet{pic98} to photometry from \citet{joh66},
\citet{cou62}, \citet{arg66}, \citet{mer86}, \citet{hau98},
\citet{ols93}, \citet{cut03}, \citet{mcc81}, and \citet{gol72}. We
furthermore use the distance calculated in \citet{van07} and set
interstellar reddening to zero.

From the SED fit, we calculate the value of HD38529's stellar
bolometric flux to be $F_{\rm BOL} = (12.02 \pm
0.084)\times10^{-8}$~erg cm$^{-2}$ s$^{-1}$ and, consequently, its
luminosity $L = 5.777 \pm 0.186 L_{\odot}$. The calculated effective 
temperature for HD~38529 is $T_{\rm eff} = 5576 \pm 74$~K (see 
Table \ref{stellar}).

%%%%%%%%%%%%%%%%%%%%%%%%%%%%%%%%%%%%%%%%%%%%%%%%%%%%%%%%%%%%%%%%%%%%

\section{Refining the Planetary Orbits}

Here we present the new RV data and the revised orbital solution for
the HD~38529 system. We combine this with the derived host star
properties to determine an accurate transit ephemeris for HD~38529b.

%%%%%%%%%%%%%%%%%%%%%%%%%%%%%%%%%%%%%%%%%%%%%%%%%%%%%%%%%%%%%%%%%%%%

\subsection{Spectra Acquisition}
\label{sec:spectra}

The RV data for HD~38529 presented here comprise 436 measurements and
were acquired from 3 instruments/telescopes: the High Resolution
Spectrograph \citep{tul98} on the Hobby-Eberly Telescope (HET), the
Hamilton Echelle Spectrograph \citep{vog87} on the 3.0m Shane
Telescope at Lick Observatory, and the HIRES echelle spectrometer
\citep{vog94} on the 10.0m Keck I telescope. Shown in Table \ref{rvs}
are a subset of the full dataset, available in the electronic version
of this paper. The fourth column in Table \ref{rvs} indicates the six
independent datasets that form the combined data, two each from HET,
Lick, and Keck. The data sources are as follows: $1 =$ HET data from
\citet{ben10}; $2 =$ new HET data presented here; $3 =$ Keck data from
\citet{wri09b}; $4 =$ Keck data from \citet{wri09b} before 2009
September 15, new Keck data presented here thereafter; $5 =$ Lick data
from \citet{wri09b}; $6 =$ Lick data from \citet{wri09b} before 2009
September 15, new Lick data presented here thereafter. The division of
the Lick data into two separate datasets is necessitated by a change
in the dewar resulting in different CCD response characteristics. The
offsets between these datasets are accounted for in the Keplerian
orbital fitting described below.

\begin{deluxetable}{cccc}
  \tablewidth{0pc}
  \tablecaption{\label{rvs} HD~38529 Radial Velocities$^\dagger$}
  \tablehead{
    \colhead{Date} &
    \colhead{RV} &
    \colhead{$\sigma$} &
    \colhead{Dataset$^\ddagger$} \\
    \colhead{(JD -- 2440000)} &
    \colhead{(m\,s$^{-1}$)} &
    \colhead{(m\,s$^{-1}$)} &
    \colhead{}
  }
  \startdata
13341.779899 & -105.27 &  7.77 & 1 \\
13341.898484 & -118.43 &  7.25 & 1 \\
13355.845730 & -102.05 &  7.34 & 1 \\
13357.859630 & -105.27 &  7.48 & 1 \\
13358.724097 &  -87.82 &  7.11 & 1 \\
13359.729188 &  -82.07 &  8.70 & 1 \\
13360.849520 &  -65.85 &  7.80 & 1 \\
13365.817387 &    1.45 &  7.73 & 1 \\
13367.812640 &  -20.41 &  9.48 & 1 \\
15095.967984 &   33.82 &  6.08 & 2 \\
15115.906428 &  -45.13 &  6.09 & 2 \\
15141.862009 &   34.65 &  6.05 & 2 \\
15142.942977 &    2.05 &  5.64 & 2 \\
15175.755713 &  -66.65 &  5.90 & 2 \\
15175.758488 &  -68.20 &  5.60 & 2 \\
15176.860583 &  -44.69 &  5.07 & 2 \\
15182.725710 &   38.24 &  4.68 & 2 \\
15185.724484 &  -15.63 &  5.29 & 2 \\
10418.959317 &   75.60 &  1.32 & 3 \\
10545.771238 &   12.67 &  1.35 & 3 \\
10787.014317 &  -69.45 &  1.40 & 3 \\
10807.061991 &  -70.45 &  1.30 & 3 \\
10837.758229 & -111.32 &  1.50 & 3 \\
10838.784387 & -113.34 &  1.43 & 3 \\
10861.729653 &  -30.86 &  1.42 & 3 \\
10862.725174 &  -41.86 &  1.46 & 3 \\
11073.058843 &  -64.92 &  1.32 & 3 \\
13750.837303 &   32.41 &  1.18 & 4 \\
13750.837905 &   32.48 &  1.32 & 4 \\
13750.839132 &   27.87 &  1.02 & 4 \\
13751.878079 &   49.46 &  0.99 & 4 \\
13752.884016 &   47.48 &  0.92 & 4 \\
13753.895081 &   33.16 &  0.98 & 4 \\
13775.747373 &   -5.19 &  0.91 & 4 \\
13776.880150 &   10.45 &  0.87 & 4 \\
14336.130035 &  261.65 &  1.11 & 4 \\
11101.015625 & -101.58 &  6.57 & 5 \\
11101.035156 & -111.09 &  6.40 & 5 \\
11102.014648 &  -88.30 &  6.17 & 5 \\
11102.032227 &  -87.47 &  6.68 & 5 \\
11131.912109 &  -91.36 &  5.99 & 5 \\
11131.930664 &  -79.77 &  7.06 & 5 \\
11132.928711 &  -67.59 &  6.02 & 5 \\
11154.808594 & -140.06 &  6.89 & 5 \\
11173.906250 & -104.89 & 10.69 & 5 \\
12267.809570 &  143.48 &  3.88 & 6 \\
12267.818359 &  150.79 &  3.86 & 6 \\
12298.721680 &  103.71 &  4.99 & 6 \\
12298.751953 &  113.82 &  5.33 & 6 \\
12298.785156 &  105.67 &  5.06 & 6 \\
12298.815430 &   93.78 &  6.53 & 6 \\
12299.695312 &  110.60 &  4.16 & 6 \\
12335.657227 &  206.29 &  4.74 & 6 \\
12534.970703 &   31.96 &  4.15 & 6 \\
  \enddata
  \tablenotetext{$\dagger$}{Shown here is a subset of the data. The
    complete dataset contains 436 measurements and is available
    electronically.}
  \tablenotetext{$\ddagger$}{Datasets 1 and 2, 3 and 4, 5 and 6 are from
    telescopes HET, Keck, and Lick respectively.}
\end{deluxetable}

%%%%%%%%%%%%%%%%%%%%%%%%%%%%%%%%%%%%%%%%%%%%%%%%%%%%%%%%%%%%%%%%%%%%

\subsection{SME Analysis}
\label{sme}

For additional insight into the properties of the host star, we used 
Spectroscopy Made Easy \citep{val96} to fit high-resolution Keck spectra of 
HD~38529. The methodology of this technique, including application of the 
wavelength intervals and line data, are described in more detail by 
\citet{val05}. We further constrained the surface gravity using Yonsei-Yale 
(Y$^2$) stellar structure models \citep{dem04} and revised \textit{Hipparcos} 
parallaxes \citep{van07} with the iterative method of \citet{val09}. The 
resulting stellar parameters are listed in Table \ref{stellar} along with 
the directly measured parameters from \S \ref{sec:fundamental_parameters}. 
The SME derived parameters are effective temperature, surface gravity, iron 
abundance, projected rotational velocity, mass, radius, and age.  These 
properties are consistent with a slightly metal-rich, mid-G sub-giant. 
There are other literature sources that have analyzed the abundances in 
HD~38529, namely: \citet{gon01}, \citet{zha02}, \citet{boda03}, \citet{ecu06}, 
\citet{gil06}, \citet{tak05}, \citet{del10}, \citet{bru11}, \citet{kang11}, 
\citet{pet11}. These catalogs contain $[$Fe$/$H$]$ values measured
with a variety of telescopes, techniques, and stellar models, and the 
determinations span the range of 0.28--0.46~dex with a mean of 0.38~dex 
and a median of 0.40~dex.  These findings are consistent with the results 
of \citet{val05}, where the average error for $[$Fe$/$H$]$ is $\pm 0.05$~dex.

The stellar radius is an essential parameter for estimating the depth and 
duration of a planetary transit. Although the spectroscopic and 
interferometric derived temeperatures are consistent with each other, the 
radii are not. This difference is a result of the SME luminosity, which 
uses bolometric corrections. These corrections can sometimes be unreliable 
for evolved stars such as HD~38529 and, in this case, is underestimated 
compared with the luminosity derived from the SED fit and parallax. Since 
a similar $T_{\rm eff}$ is found from the two techniques, this results in 
a reduced radius estimate. As described in \S \ref{ephem}, we adopt the 
interferometric radius for the purposes of estimating the predicted transit 
properties.

%%%%%%%%%%%%%%%%%%%%%%%%%%%%%%%%%%%%%%%%%%%%%%%%%%%%%%%%%%%%%%%%%%%%

\subsection{Keplerian Orbital Solution}

\begin{figure}
  \includegraphics[width=8.2cm]{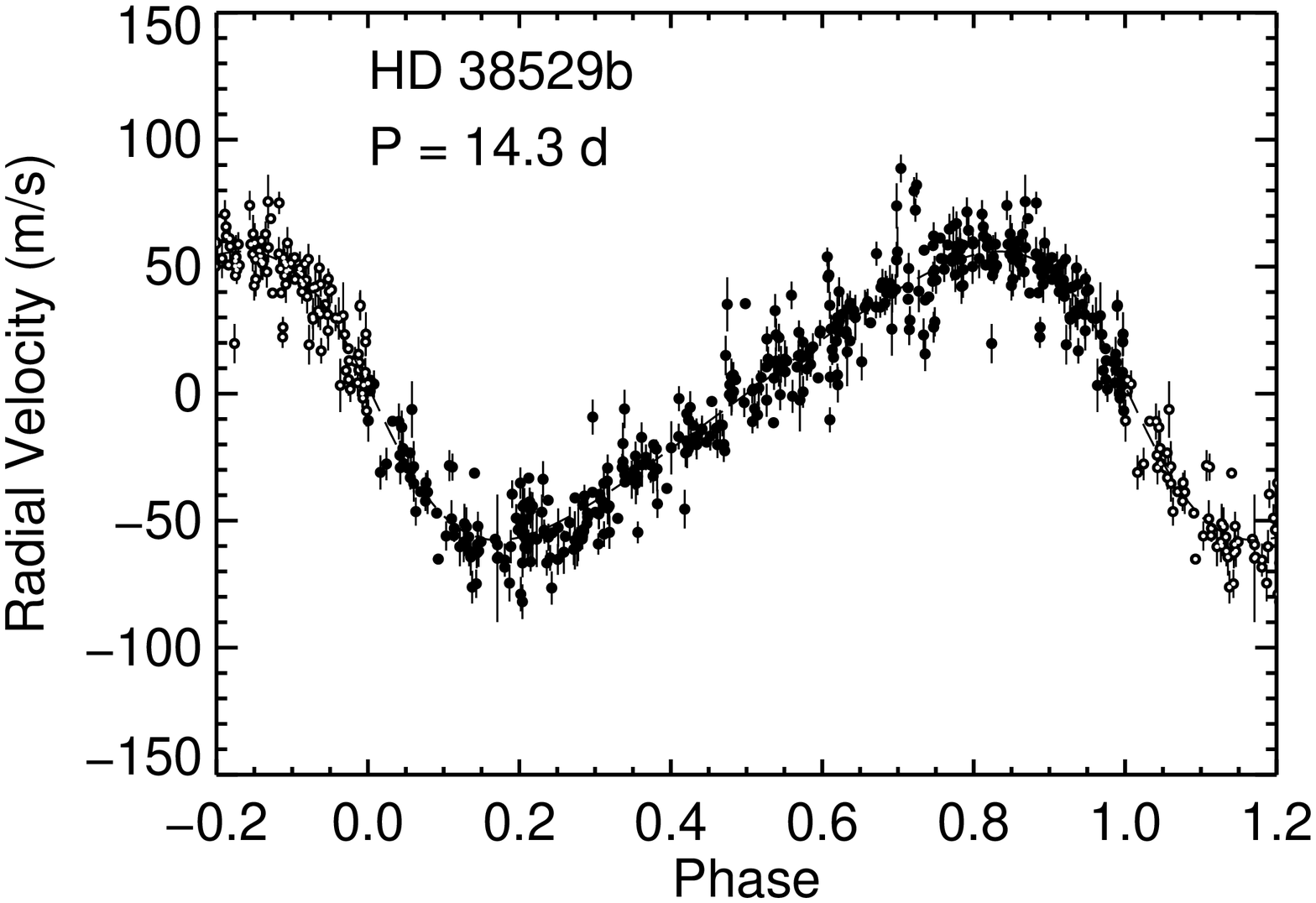}
  \includegraphics[width=8.2cm]{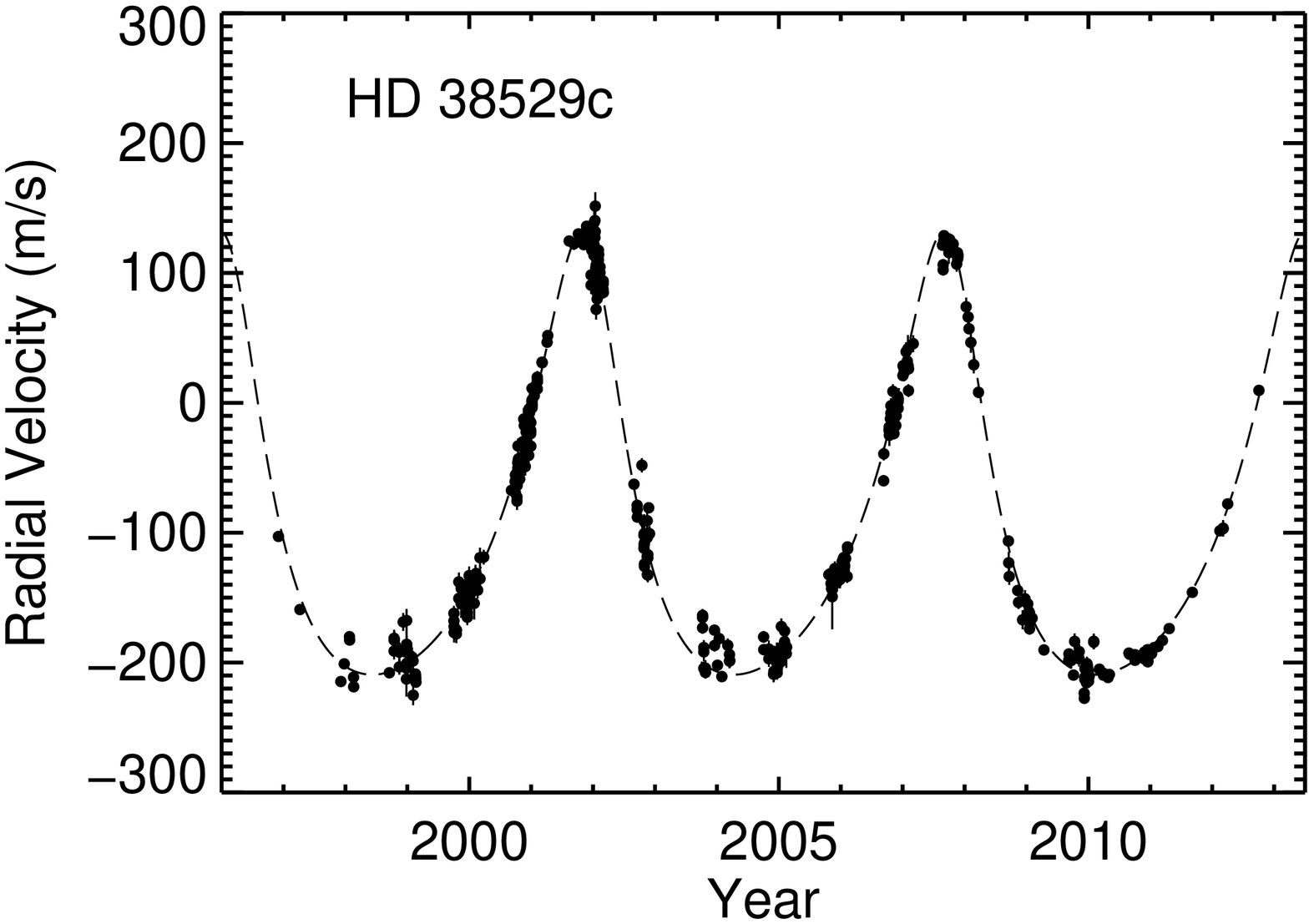}
  \includegraphics[width=8.2cm]{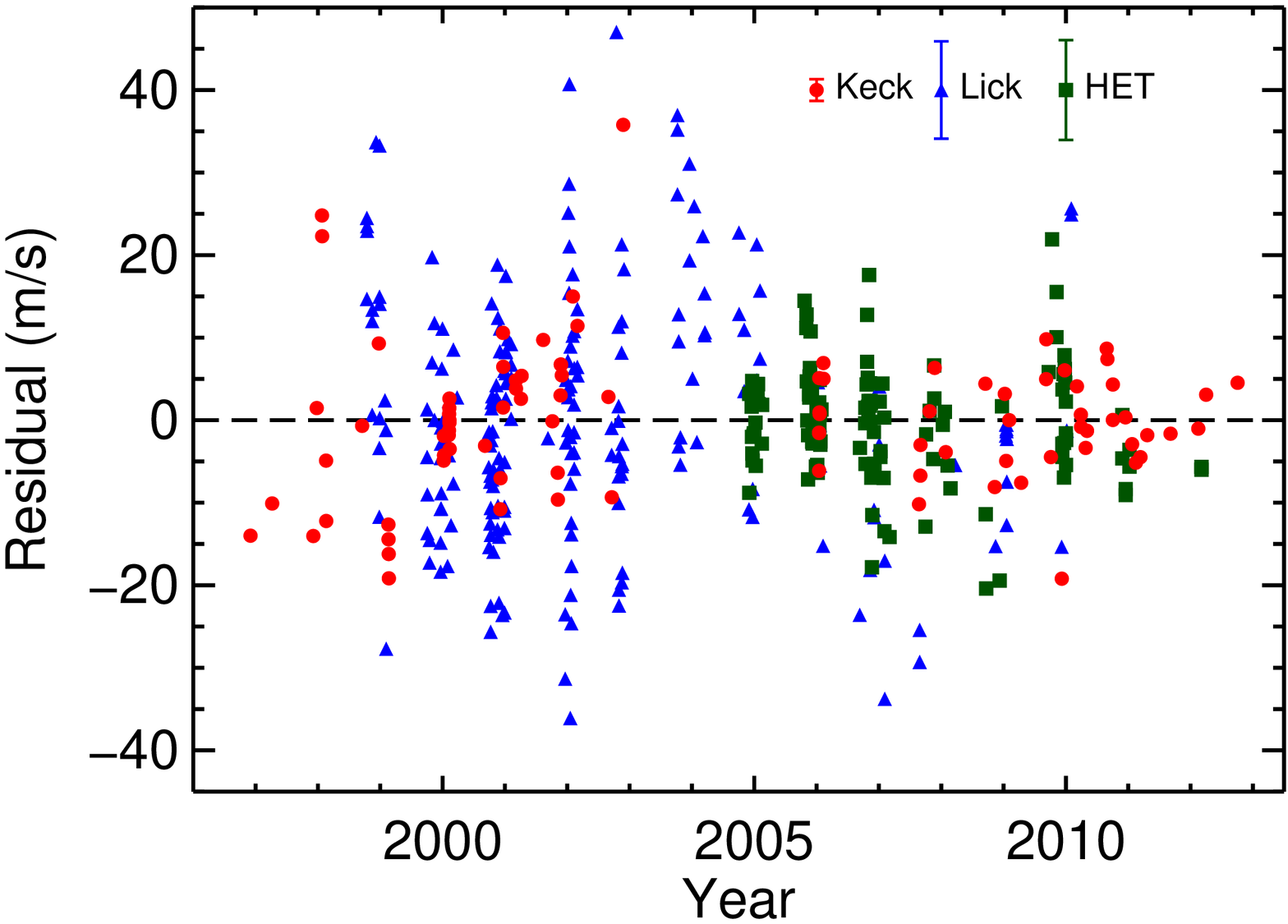}
  \caption{Top and Middle Panels: Radial velocity signal (black dots)
    induced by HD 38529b and c, respectively, and the best-fit
    orbital solution (dashed line). Error bars shown are internal
    errors for each observation. The radial velocity signal for each
    planet was extracted by subtracting off the best-fit orbital
    velocities of the other planet from the total observed RVs. Bottom
    Panel: Residual velocities with respect to the best 2-planet
    orbital solution. The red dots are for Keck data (dataset 3 and
    4), the blue triangles are for Lick data (dataset 5 and 6), and
    the green squares are for the HET data (dataset 1 and 2). The
    typical size of internal error bars for each telescope ($\pm$
    median internal errors) are plotted on the upper right of this
    panel.}
  \label{rv}
\end{figure}

We fit a two-planet Keplerian orbital solution to the RV data using
the partially linearized, least-squares fitting procedure described in
\citet{wri09a} and estimated parameter uncertainties using the
BOOTTRAN bootstrapping routines described in \citet{wan12}. Table
\ref{planet} lists the resulting parameters for the two-planet orbital
solution. The data and orbital solutions are shown separately for the
b and c planets in Figure \ref{rv}, along with the residual velocities
with respect to the best 2-planet orbital solution. The improved mass
estimate for planet c combined with the Fine Guidance Sensor (FGS)
astrometry of \citet{ben10} yields a true mass estimate of $16.76 \pm
0.11$~$M_J$. As noted by \citet{ben10}, this lies within the brown
dwarf mass regime.

\begin{deluxetable}{lc}
  \tablecaption{\label{planet} Keplerian Orbital Model}
  \tablewidth{0pt}
  \tablehead{
    \colhead{Parameter} &
    \colhead{Value}
  }
  \startdata
\noalign{\vskip -3pt}
\sidehead{HD 38529 b}
~~~~$P$ (days)                    & $14.30978 \pm 0.00033$ \\
~~~~$T_c\,^{a}$ (JD -- 2,440,000) & $15815.633 \pm 0.063$ \\
~~~~$T_p\,^{b}$ (JD -- 2,440,000) & $12281.19 \pm 0.15$ \\
~~~~$e$                           & $0.259 \pm 0.016$ \\
~~~~$K$ (m\,s$^{-1}$)             & $56.81 \pm 1.01$ \\
~~~~$\omega$ (deg)                & $93.3 \pm 4.1$ \\
~~~~$M_p$\,sin\,$i$ ($M_J$)       & $0.8047 \pm 0.0139$ \\
~~~~$a$ (AU)                      & $0.1278 \pm 0.0006$ \\
\sidehead{HD 38529 c}
~~~~$P$ (days)                    & $2133.54 \pm 3.31$ \\
~~~~$T_p\,^{b}$ (JD -- 2,440,000) & $12264.49 \pm 6.43$ \\
~~~~$e$                           & $0.3472 \pm 0.0057$ \\
~~~~$K$ (m\,s$^{-1}$)             & $170.54 \pm 1.12$ \\
~~~~$\omega$ (deg)                & $20.08 \pm 1.14$ \\
~~~~$M_p$\,sin\,$i$ ($M_J$)       & $12.51 \pm 0.08$ \\
~~~~$a$ (AU)                      & $3.594 \pm 0.018$ \\
\sidehead{System Properties}
~~~~$\gamma$ (m\,s$^{-1}$)           & $96.34 \pm 3.59$ \\
\sidehead{Measurements and Model}
~~~~$N_{\mathrm{obs}}$            & 436 \\
~~~~rms (m\,s$^{-1}$)             & 11.76 \\
~~~~$\chi^2_{\mathrm{red}}$       & 11.67 \\
  \enddata
  \tablenotetext{a}{Time of mid-transit.}
  \tablenotetext{b}{Time of periastron passage.}
\end{deluxetable}

The fit required five additional free parameters due to the offsets
between the six independent datasets. The offsets with respect to the
observations acquired with HET dataset 2 (the new HET data presented
in this paper, see \S \ref{sec:spectra}) are $-44.9$, $-63.92$,
$-73.01$, $-117.2$, and $-123.1$ m\,s$^{-1}$ for datasets 1, 3, 4, 5,
and 6 respectively. As shown in Table \ref{planet}, the
$\chi^2_{\mathrm{red}}$ and rms scatter of the residuals are
relatively large. The major contributor to the calculated
$\chi^2_{\mathrm{red}}$ value are the Keck data whose uncertainties
are significantly smaller than those associated with the Lick and HET
data.  Specifically, the Keck rms scatter of 8.6 m\,s$^{-1}$ exceeds
its internal error, which has a median value of 1.3 m\,s$^{-1}$. This
indicates that there is a stellar noise component to the overall noise
level which is not accounted for in the fit. The causes of the stellar
noise, including pulsations and star spot activity, are discussed in
detail by \citet{ben10}.  Another hypothesis is that of an additional
planet with an amplitude of $K \gtrsim 5$~m\,s$^{-1}$. We explain in
the following section that our data do not support detection of such
an additional planet.

%%%%%%%%%%%%%%%%%%%%%%%%%%%%%%%%%%%%%%%%%%%%%%%%%%%%%%%%%%%%%%%%%%%%

\subsection{A Third Planet?}
\label{thirdplanet}

\citet{ben10} utilize their results to speculate on evidence for a third 
planet in the system. Thus, we also consider this possibility from our 
analysis since our RV data comprise a substantially larger dataset. As 
reported by Benedict et al. (2010), a coplanar orbital solution is only 
stable if the third planet has a period within the window of $[33,445]$ days 
and an eccentricity of $e <0.3$, or a period larger than their RV data 
baseline ($>10$ years). For this reason, we focused our search for the 
third planet within the period window of $[33,445]$ days and constrained 
the eccentricity to be $<0.3$.

\begin{figure}
  \includegraphics[width=8.2cm]{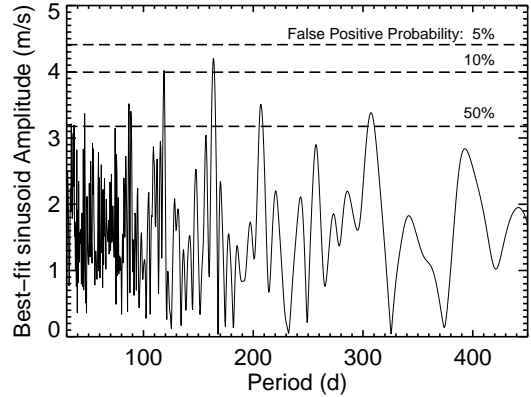}
  \caption{Amplitude of best-fit sinusoids to the residuals of the
    two-planet Keplerian solution (solid line). Any peak in this
    period window that has amplitude larger than the top dashed line
    is considered to be significant for having $<5\%$ false positive
    probability. The two lower dashed lines ($<10\%$ and $<50\%$) have 
    similar meanings. No period within this window has less than
    5\% false positive probability, and the two peaks with $<10\%$
    false positive probability are at 119 days and 164 days. We see no
    significant peak around 194 days as reported by \citet{ben10}. See
    Section~3.4 for more details.}
  \label{pgram}
\end{figure}

We first searched for strong periodic signals in the residuals of the 
two-planet Keplerian solution by fitting sinusoids to the residuals at 
different periods within $[33,445]$ days (with 0.4 day step in period). 
The results are plotted as the solid line in Figure~\ref{pgram}. We then 
estimated the false positive probability to see if any of the strong peaks 
are significant. We define the false positive probability for a peak with a 
certain amplitude $K'$ as the probability that a signal with amplitude 
$\geq K'$ is generated by the residuals by chance. We generated 1000 sets 
of simulated residuals by scrambling the true residuals (and their 
associated errors, with replacements), and then searched for the peak 
with largest amplitude within the $P=[33,445]$ day window for each of the 
1000 scrambled data sets. These 1000 amplitudes provide approximately the 
distribution of amplitudes arising purely from random noise in the 
residuals. Any peak in Figure~\ref{pgram} that has an amplitude smaller 
than 950 (95\%) of these 1000 amplitudes is thus considered having false 
positive probability of $>5\%$. This is marked by the top dashed line in 
Figure~\ref{pgram} and similarly for the $10\%$ and $50\%$ lines.

As shown in Figure~\ref{pgram}, no peak has a false positive probability 
of less than 5\%, and there are just two with less than 10\% at 119 days 
and 164 days. The highest peak at 164 days has a false positive probability 
of 6.8\%. We see no significant peak around 194 days as reported by 
\citet{ben10}. We then performed 3-planet Keplerian fit with our RV data 
within the $P=[33,445]$ day window and with the constraint that the 
eccentricity must be smaller than 0.3. We found that indeed the best-fit 
is near 164 days, with $e=0.3$ (also true if we force the third planet to 
be on a circular orbit; best-fit $e=0.99$ if no constraint on $e$ is 
required). The $\chi_{\nu}^2$ of this fit is 9.58, and an f-test suggests 
that the 3-planet model provides a better fit though having 5 more 
parameters.  However, the rms for this fit is 11.92 m s$^{-1}$, i.e., 
adding a third planet does not reduce the rms of the fit. Combining with 
the fact that this signal at $P=164$ days does not have lower than 5\% 
false positive probability, we cannot conclude that our data have detected 
a third planet in the HD~38529 system.

We note here that including or excluding this third planet does not
affect our transit exclusion analysis in the following sections,
because the changes in the orbital parameters for both HD 38529b or
c, after adding the third planet, are smaller than their error bars
listed in Table~\ref{planet}.

%%%%%%%%%%%%%%%%%%%%%%%%%%%%%%%%%%%%%%%%%%%%%%%%%%%%%%%%%%%%%%%%%%%%

\subsection{Transit Ephemeris Refinement}
\label{ephem}

From the stellar and planetary properties listed in Tables \ref{stellar} 
and \ref{planet}, we compute a refined transit ephemeris for HD~38529b. We 
use the directly measured radius of the host star determined from our 
CHARA observations. We approximate the radius of the planet as 
$R_p = 1.0 R_J$, based upon the mass-radius relationship described by 
\citet{kan12}. These properties lead to a predicted transit duration of 
0.33 days and a predicted transit depth of 0.17\% or 0.0018 mag. 
\citet{kan08} show how the probability of a planetary transit is a strong 
function of both the eccentricity and the argument of periastron. The 
periastron argument of HD~38529 is particularly well-aligned for transit 
probability enhancement since it is quite close to the optimal angle of 
$90\degr$. Combined with the eccentricity, this results in a transit 
probability of 12.8\%. However, if the planet were in a circular orbit, 
the duration would be increased to 0.43 days but the probability would be 
reduced to 9.5\%.

The calculation of the transit mid-point shown in Table \ref{planet} was 
performed with a Monte-Carlo bootstrap, which propagates the uncertainty 
in this orbital parameter to the time of the transit. The resulting 
uncertainty in the transit mid-point is 
$0.063\mathrm{ days} = 91 \mathrm{ minutes}$. As such, the transit window 
is dominated by the predicted duration rather than the mid-point uncertainty, 
making it a suitable candidate for photometric follow-up \citep{kan09}. 
Thus, we compute a revised transit ephemeris from the orbital fit, which we 
utilize in the following sections.

%%%%%%%%%%%%%%%%%%%%%%%%%%%%%%%%%%%%%%%%%%%%%%%%%%%%%%%%%%%%%%%%%%%%

\section{Photometric Observations}

We have acquired 1106 photometric observations of HD~38529 on 992
nights between 2000 November 28 and 2012 March 31, all with the T11
0.80~m Automated Photoelectric Telescope (APT) at Fairborn Observatory
in Arizona. The T11 APT, one of several such telescopes operated at
Fairborn by Tennessee State University, is equipped with a two-channel
precision photometer that uses a dichroic filter and two EMI 9124QB
bi-alkali photomultiplier tubes to separate and simultaneously measure
the Str\"omgren $b$ and $y$ pass bands. We programmed the APT to make
differential brightness measurements of our program star HD~38529 (P,
$V=5.95$, $B-V=0.77$, G4~IV) with respect to the two comparison stars
HD~38145 (C1, $V=7.89$, $B-V=0.33$, F0~V) and HD~40259 (C2, $V=7.86$,
$B-V=0.38$, F0~V). From the raw counts in both pass bands, we compute
the differential magnitudes $P-C1$, $P-C2$, and $C2-C1$, correct them
for atmospheric extinction, and transform them to the Str\"omgren
system. To improve our photometric precision, we combine the
differential $b$ and $y$ observations into a single $(b+y)/2$
passband, which we indicate with the subscript $by$. Furthermore, we
compute the differential magnitudes of HD~38529 against the mean
brightness of the two comparison stars. The resulting precision of
the individual $P-(C1+C2)/2_{by}$ differential magnitudes ranges
between $\sim0.0010$ mag and $\sim0.0015$ mag on good nights, as
determined from the nightly scatter in the $C2-C1$ differential
magnitudes of the constant comparison stars. Further details of our
automatic telescopes, precision photometers, and observing and data
reduction procedures can be found in \citet{h1999} and \citet{ehf2003}
and references therein.

\begin{figure}
  \includegraphics[width=8.2cm]{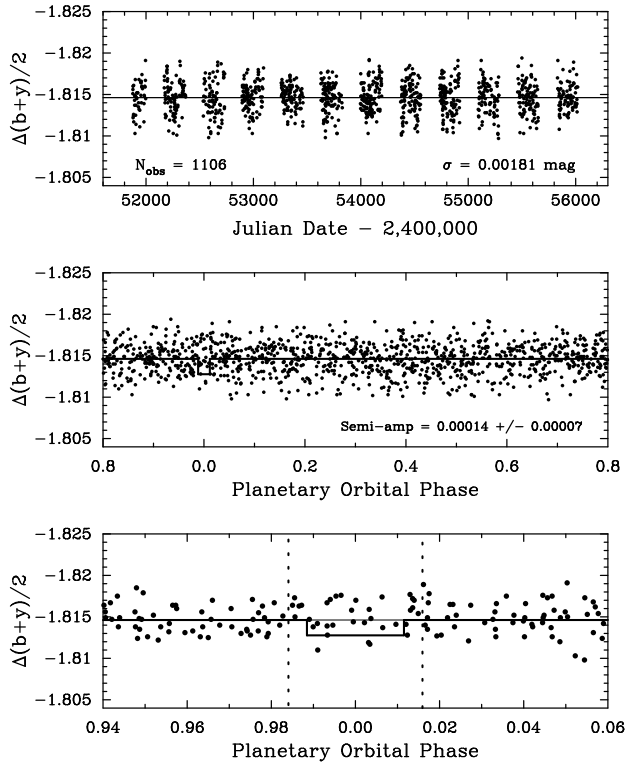}
  \caption{$Top$: The 1106 individual $P-(C1+C2)/2_{by}$ differential
    magnitudes of HD~38529, acquired with the T11 0.8~m APT during 12
    observing seasons between 2000 and 2012. The observations in each
    observing season have been normalized to give each observing
    season the same yearly mean. $Middle$: The 1106 observations
    phased with the orbital period and time of transit of companion b.
    The semi-amplitude of a least-squares sine fit to this orbital
    phase curve is $0.00014~\pm~0.00007$~mag, consistent with the
    absense of periodic light variability in HD~38529 on the radial
    velocity period. This provides strong confirmation of the
    existence of planet b. $Bottom$: The observations near phase 0.0
    plotted on an expanded scale. The solid curve shows the estimated
    depth ($\sim0.0018$ mag) and duration ($\pm 0.012$ phase units) of
    a central transit of companion b. The $\pm1\sigma$ extent of the
    transit window is indicated by the vertical dotted lines. Transits
    are ruled out to a depth of $\sim0.0004$ mag.}
  \label{photfig1}
\end{figure}

The resulting 1106 differential magnitudes, spanning 12 consecutive
observing seasons, are summarized in Table~\ref{phottab} and plotted
in the top panel of Figure~\ref{photfig1}, after normalization to
bring the seasonal means to a common level indicated by the
horizonatal line in the top panel. The normalization removes
season-to-season variability in HD~38529 perhaps caused by a starspot
cycle in this mildly active star (see below).  The scatter in the
normalized data from their grand mean is $\sigma~=~0.0019$ mag
(standard deviation).  This is slightly larger than our typical
measurement precision given above, suggesting low-amplitude,
night-to-night variability in HD~38529.

\begin{deluxetable*}{ccccccccc}
  \tablewidth{0pt}
  \tablecaption{\label{phottab} Summary of Photometric Observations
    for HD~38529}
  \tablehead{
    \colhead{Observing} & \colhead{} & \colhead{Julian Date Range} &
    \colhead{Sigma} & \colhead{$P_{rot}$} & \colhead{Full Amplitude} &
    \colhead{$<P-C1>$} & \colhead{$<P-C2>$} & \colhead{$<C2-C1>$} \\
    \colhead{Season} & \colhead{$N_{obs}$} & \colhead{(HJD $-$ 2,400,000)} &
    \colhead{(mag)} & \colhead{(days)} & \colhead{(mag)} &
    \colhead{(mag)} & \colhead{(mag)} & \colhead{(mag)} \\
    \colhead{(1)} & \colhead{(2)} & \colhead{(3)} &
    \colhead{(4)} & \colhead{(5)} & \colhead{(6)} &
    \colhead{(7)} & \colhead{(8)} & \colhead{(9)}
  }
  \startdata
  2000--01 &  47 & 51877--51997 & 0.00181 &    \nodata      &      \nodata   & $-1.8254\pm0.0002$ & $-1.8033\pm0.0002$ & $-0.0221\pm0.0001$ \\
  2001--02 &  88 & 52169--52368 & 0.00190 & $38.5\pm0.5$ & $0.0024\pm0.0005$ & $-1.8288\pm0.0002$ & $-1.8061\pm0.0002$ & $-0.0226\pm0.0001$ \\
  2002--03 &  76 & 52532--52732 & 0.00219 & $36.3\pm0.2$\tablenotemark{a} & $0.0042\pm0.0007$ & $-1.8303\pm0.0002$ & $-1.8076\pm0.0002$ & $-0.0226\pm0.0001$ \\
  2003--04 &  95 & 52894--53094 & 0.00166 & $36.2\pm0.3$ & $0.0029\pm0.0004$ & $-1.8288\pm0.0002$ & $-1.8056\pm0.0001$ & $-0.0232\pm0.0001$ \\
  2004--05 & 103 & 53258--53462 & 0.00138 & $35.7\pm0.3$ & $0.0022\pm0.0003$ & $-1.8259\pm0.0001$ & $-1.8033\pm0.0001$ & $-0.0225\pm0.0001$ \\
  2005--06 & 101 & 53627--53827 & 0.00175 &    \nodata      &      \nodata   & $-1.8240\pm0.0001$ & $-1.8017\pm0.0001$ & $-0.0223\pm0.0001$ \\
  2006--07 & 118 & 53996--54194 & 0.00177 & $38.0\pm0.2$\tablenotemark{a} & $0.0026\pm0.0004$ & $-1.8246\pm0.0001$ & $-1.8015\pm0.0001$ & $-0.0231\pm0.0001$ \\
  2007--08 & 110 & 54370--54556 & 0.00187 & $37.3\pm0.2$\tablenotemark{a} & $0.0021\pm0.0005$ & $-1.8265\pm0.0002$ & $-1.8037\pm0.0001$ & $-0.0227\pm0.0001$ \\
  2008--09 & 104 & 54734--54919 & 0.00198 &    \nodata      &      \nodata   & $-1.8273\pm0.0002$ & $-1.8047\pm0.0002$ & $-0.0226\pm0.0001$ \\
  2009--10 &  86 & 55092--55284 & 0.00189 & $36.5\pm0.3$ & $0.0034\pm0.0006$ & $-1.8275\pm0.0002$ & $-1.8042\pm0.0001$ & $-0.0232\pm0.0001$ \\
  2010--11 &  95 & 55459--55650 & 0.00184 &    \nodata      &      \nodata   & $-1.8253\pm0.0001$ & $-1.8022\pm0.0002$ & $-0.0230\pm0.0001$ \\
  2011--12 &  83 & 55830--56018 & 0.00157 & $37.2\pm0.4$ & $0.0024\pm0.0004$ & $-1.8252\pm0.0002$ & $-1.8019\pm0.0001$ & $-0.0232\pm0.0001$ \\
  \enddata
  \tablenotetext{a}{Periodogram analysis gave half of the quoted
    period, implying the star had spots on both hemispheres at those
    epochs.  We doubled the photometric periods and their errors in
    these cases to get $P_{rot}$.}
\end{deluxetable*}

The normalized observations from all 12 observing seasons are
replotted in the middle panel of Figure~\ref{photfig1}, phased with
the 14.3~day planetary orbital period of HD~38529b and the time of mid
transit ($T_c$) from Table~\ref{planet}. A least-squares sinusoidal
fit to the phased data gives a formal semi-amplitude of just
$0.00014~\pm~0.00007$ mag, which limits any periodic brightness
variability of the star on the orbital period to a very small fraction
of one milli-magnitude (mmag). This rules out the possibility that the
14.3-day radial velocity variations are simply jitter induced by
stellar activity, as has been documented in slightly more active
stars, for instance, by \citet{qhs+2001}, \citet{psch2004}, and
\citet{bbs2012}. Instead, the lack of photometric variability confirms
that the radial velocity variations in HD~38529 result from true
planetary reflex motion.

The photometric observations within $\pm 0.06$ phase units of
mid-transit are plotted with an expanded scale in the lower panel of
Figure~\ref{photfig1}. The solid curve shows predicted transit phase
(0.0), depth ($\sim0.0018$ mag), and duration ($\pm 0.012$ phase
units) of a central transit, all computed from the stellar radius in
Table~\ref{stellar} and the orbital elements of HD~38529b in
Table~\ref{planet}. The vertical dotted lines give the $\pm1\sigma$
uncertainty in the timing of the transit window, based on the
uncertainties in the stellar radius and the improved orbital elements
from Tables \ref{stellar} \& \ref{planet}, respectively. Our data set
contains 1084 photometric observations that lie outside the predicted
transit time (solid curve); these have a mean of
$-1.81462~\pm~0.00005$ mag. There are 22 observations that fall in
transit; these have a mean of $-1.81474~\pm~0.00041$ mag.  The
difference is our ``observed transit depth,'' $-0.00012~\pm0.00042$
mag in the sense that the mean of the transit points is slightly 
{\it brighter} than the mean of the out-of-transit observations but are,
nevertheless, consistent with zero to four decimal places. As the
lower panel shows, we have sufficient data around predicted transit 
times to rule out transits to a depth of $\sim0.0004$ mag.

It has been suggested by \citet{ang10} that planetary systems in 2:1
orbital resonance can be mistaken for a single planet in an eccentric
orbit. If planet b has an orbital period that is half of the
derived $\sim 14.3$~day period, then it would have an even higher
transit probability and thus a higher chance of being detected in the
photometric data. We performed a search through the 12-year
photometric data set for periodic transits around 7.155 days and
around 14.3 days. No transits were detected at or near either period. 
Although a dynamical model of a 2:1 resonance would lead to aperiodic 
transit times, the complete phase coverage of our photometric observations 
over a range of periods around 14.3 days leads us to conclude that such 
transits are ruled out.

\begin{figure}
  \includegraphics[width=8.2cm]{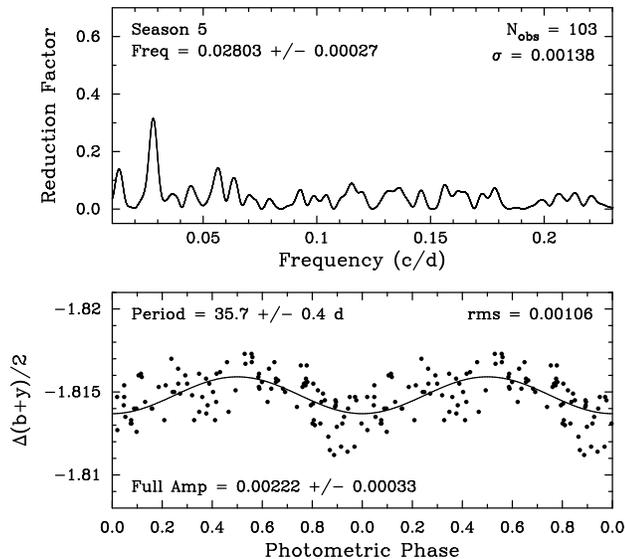}
  \caption{$Top$: A frequency spectrum of the 2004--2005 (Season~5)
    photometric observations of HD~38529. The best frequency occurs at
    $0.02803\pm0.00027$ cycles per day. $Bottom$: The 103 Season~5
    observations phased with the corresponding best period of 35.7
    days. The phase curve shows coherent variability with a
    peak-to-peak amplitude of 0.002 mag, which we take to be
    rotational modulation of photospheric spots. Eight of the twelve
    observing seasons exhibit similar modulation (see Table~\ref{phottab}).}
  \label{photfig2}
\end{figure}

As noted above, the scatter in the normalized data set is slightly larger 
than the expected measurement precision, suggesting the presence of small 
starspots on HD~38529. Starspots on the photospheres of solar-type stars 
allow the possibility of direct determination of stellar rotation periods 
from rotational modulation in the visibility of the spots and the consequent 
variability in the star's brightness \citep[see, e.g.,][]{sbh+2010}. Spots 
can also produce periodic radial velocity variations that can mimic the 
presence of a planetary companion.  Therefore, we performed periodogram 
analyses for each of the 12 seasons of normalized photometry plotted in 
the top panel of Figure~\ref{photfig1} and, indeed, found very low-amplitude 
periodic brightness variations in 8 of the 12 observing seasons. Similar 
analyses of the twelve $C2-C1$ seasonal datasets evinced no significant 
periodicity in the comparison stars. Figure~\ref{photfig2} shows a sample 
frequency spectrum and phase curve for Season 5. Complete results are given 
in columns 5 \& 6 of Table~\ref{phottab}. The seasonal photometric periods 
scatter about their mean with a standard deviation of $\sim1$ day. The 
weighted mean of the 8 photometric periods is $37.0\pm0.4$ days, which we 
take to be our best determination of the star's rotation period.  This 
period is based on far more observations than the preliminary periods of
35.7 and 31.6 days given by \citet{fis03} and \cite{ben10}, respectively. 
The seasonal peak-to-peak amplitudes in column~6 range between 0.002 to 
0.004 mag, indicating spot filling factors of only a few tenths of one 
percent. Both the low level of spottedness and the fact that the stellar 
rotation period is distinctly different from the 14-day radial velocity 
period and its harmonics demonstrates that stellar activity (spots and 
plages) is not the source of the 14-day radial velocity period.

If the inclination of HD~38529's rotation axis is near $90\arcdeg$, so that 
$V\sin~i$ approximately equals the equatorial rotation velocity, $V_{eq}$, 
then the stellar radius and projected rotational velocity from 
Table~\ref{stellar} result in a rotation period of $\sim37.5$ days, 
essentially identical to our observed value of $P_{rot}=37.0$ days. This 
implies a very high inclination for the stellar rotation axis and increases 
the probability of transits of HD~38529b, assuming the stellar equatorial 
and planetary orbital planes are aligned. \citet{mdu+2011} have shown, 
for instance, that the orbital planes of the hot jupiters HAT-P-8b, HAT-P-9b, 
HAT-P-16b, \& HAT-P-23b are all closely aligned with the stellar rotation 
axis.  In the same paper, \citet{mdu+2011} examined 37 exoplanetary systems 
that have accurately measured spin-orbit angles and found that spin-orbit 
misalignment occurs primarily for stars with $T_{eff}~>~6300$~K (spectral 
class F6 and hotter). Thus, the orbital geometry, the high stellar 
inclination, and the likelihood of spin-orbit alignment are all favorable 
for transits of HD~38529b, making our non-detection particularly 
disappointing.

\begin{figure}
  \includegraphics[width=8.2cm]{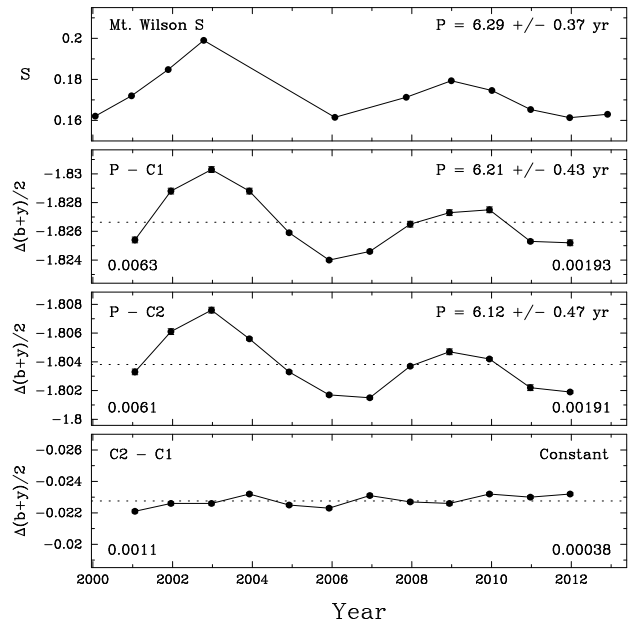}
  \caption{Long-term variations in Mt. Wilson S-index ($Top$) and
    brightness ({\it Panels 2 and 3}) measured with Keck I and the T11
    APT. The cycle timescale is $\sim 6$~yrs in all three timeseries.
    Long-term differential magnitudes of the two comparison stars C1
    and C2 ($Bottom$) show excellent stability to $\pm0.00038$ mag),
    demonstrating that the variability in the $P-C1$ and $P-C2$ light
    curves is intrinsic to HD~38529.  The direct correlation of
    S-index and brightness is typical for stars of solar age and
    older. The weak magnetic activity in HD~38529 cannot produce the
    large-amplitude radial velocity variations attributed to planet
    c.}
  \label{photfig3}
\end{figure}

Finally, we examine long-term variability in HD~38529's Ca~II~H~and~K
indices and APT photometry to look for evidence of magnetic cycles
that might induce apparent radial velocity variations and so mimic the
presence of a {\it long-period} planet. Both H~and~K emission and
brightness variability are good proxies for stellar magnetic activity
\citep[see, e.g.,][and references
  therein]{bal+1995,bdsh1998,lock+2007}.  In the top panel of
Figure~\ref{photfig3}, we plot seasonal means of the Mount Wilson
S-index derived from our Keck I RV spectra as described in
\citet{wri04} and \citet{if2010}. Despite a couple of observing
seasons without Keck H~and~K measurements, clear variability is seen
on a timescale of several years. In panels 2 \& 3, we plot the 12
seasonal mean $P-C1$ and $P-C2$ differential magnitudes
(Table~\ref{phottab}) without any normalization, as was applied to the
$P-(C1+C2)/2$ differential magnitudes in Figure~\ref{photfig1}. These
two lightcurves show that HD~38529 varies in brightness by several
mmag from year-to-year with respect to both comparison stars C1 and
C2. The bottom panel of Figure~\ref{photfig3} plots the yearly mean
$C2-C1$ comparison star differential magnitudes. The number in the
lower left corner of each panel gives the range of the seasonal means
in magnitudes; the number in the lower right corner gives the standard
deviation of the seasonal means with respect to their grand mean,
shown by the horizontal dotted line in each panel.  The standard
deviation of the $C2-C1$ observations is only 0.00038 mag, indicating
that both comparison stars have excellent long-term photometric
stability.  Therefore, the short- and long-term variability evident in
the $P-C1$ and $P-C2$ light curves must be intrinsic to HD~38529.

We see cyclic variation in both the H~and~K and APT observations plotted 
in Figure \ref{photfig3}. Simple least-squares, sine-fit periodogram 
analyses of the H~and~K, $P-C1$, and $P-C2$ means give periods of 
$6.29\pm0.37$, $6.21\pm0.43$, and $6.12~\pm~0.47$~years, respectively, 
all indentical within their uncertainties. Futhermore, the brightness of 
the star and the strength of the H~and~K emission vary directly in phase 
with each other. The cycle timescale and amplitudes of the H~and~K and 
photometric observations are typical of similar long-term cycles seen in 
a large sample of solar-type stars being measured with the APTs 
\citep[see][]{h1999,lock+2007,hhl+2009} and are analogous to solar 
irradiance variations measured for over three decades with space-based 
radiometers \citep[see, e.g.,][]{wh1981,f2012}. The direct correlation of 
long-term brightness and magnetic activity in HD~38529 is in the same 
sense as the correlation between solar irradiance and S-index. Despite 
the similarities in age and spectral class of the Sun (4.6~Gyr, G2) and 
HD~38529 (4.4~Gyr, G5), the amplitudes of the S-index and brightness 
variability are larger in HD~38529 by factors of approximately 2 and 5, 
respectively \citep[see, e.g.][their Fig. 2]{lock+2007}. This is due to
the fact that HD~38529 is a more massive, evolved subgiant (Table~1) 
with a deeper convection zone.

The radial velocity period of HD~38529c is 2134.9 days (Table~\ref{planet}) 
or 5.85 years, very close to the spot-cycle timescale of 6.2 years. This 
raises the question that the radial velocity variations attributed to 
HD~38529c may, instead, originate from the stellar magnetic cycle. As 
mentioned above, starspots in active stars have been found to induce 
radial velocity variations on stellar rotation timescales. Over the past 
15 years or so, several authors have discussed the mechanisms by which 
long-term (decadal) magnetic cycles might be the source of long-term 
radial velocity variations \citep[e.g.,][]{sd1997,sf2000,sslm2010}. 
Recently, \citet{lds+2011} analyzed a sample of 304 FGK stars from the 
HARPS high-precision planet-search sample, looking for possible correlations 
of stellar activity cycles with radial velocity and spectral line-shape 
parameters. They confirmed that stellar magnetic cycles {\it can} induce 
long-period, low-amplitude radial velocity variations with amplitudes up 
to 25~m\,s$^{-1}$ in the more active stars. For HD~38529c, however, its 
radial velocity semi-amplitude of 171~m\,s$^{-1}$ makes it immune to this 
type of false positive.

%%%%%%%%%%%%%%%%%%%%%%%%%%%%%%%%%%%%%%%%%%%%%%%%%%%%%%%%%%%%%%%%%%%%

\section{Conclusions}

With the discovery of extrasolar planets, characterization has become
an important pursuit and is intrinsically linked to the understanding
of the host star properties. Thus, even relatively bright stars, such
as HD~38529, are worthy of further scrutiny in order to gain insight
into properties of the planets hosted by the star. Here we have
performed just such a task by presenting the most accurate properties
of the star HD~38529 so far produced. Direct measurements of stellar
radii via interferometry provide an essential test of stellar models
and validation of radii derived from spectroscopy. In this case, we
measure a radius of 2.578~R$_\odot$ for HD~38529, which is consistent
with a slightly metal-rich mid-G sub-giant. The new RV data for the
star are used to calculate a new Keplerian orbital solution for the
system that enables us to place significant limits on a previously
postulated third planet. Combining our mass estimate for the c
component with the FGS astrometry of \citet{ben10} allows us to
confirm that this object does lie within the brown dwarf mass regime.
The refined transit ephemeris from the Keplerian orbital solution is
combined with 12 years of precision photometry to demonstrate the
variability of the host star and the dispositive null detection of
transits for HD~38529b. The importance of this study is clear since
the combination and inter-dependence of all these effects leads to a
greatly improved understanding of the system as a whole. The TERMS
project is systematically proceeding to provide such characterization
of the planetary systems known to orbit the brightest stars in the
sky.

%%%%%%%%%%%%%%%%%%%%%%%%%%%%%%%%%%%%%%%%%%%%%%%%%%%%%%%%%%%%%%%%%%%%

\section*{Acknowledgements}

The Center for Exoplanets and Habitable Worlds is supported by the
Pennsylvania State University, the Eberly College of Science, and the
Pennsylvania Space Grant Consortium. S.R.K and N.R.H. acknowledge
financial support from the National Science Foundation through grant
AST-1109662. G.W.H. acknowledges support from NASA, NSF, Tennessee
State University, and the State of Tennessee through its Centers of
Excellence program. The HET is a joint project of the University of
Texas at Austin, the Pennsylvania State University, Stanford
University, Ludwig-Maximilians-Universit\"at M\"unchen, and
Georg-August-Universit\"at G\"ottingen. The HET is named in honor of
its principal benefactors, William P. Hobby and Robert E. Eberly.

%%%%%%%%%%%%%%%%%%%%%%%%%%%%%%%%%%%%%%%%%%%%%%%%%%%%%%%%%%%%%%%%%%%%

\end{document}